\documentclass[twocolumn,prb,english,showpacs]{revtex4-1}
\usepackage[latin9]{inputenc}
\usepackage{amstext,bm}
\usepackage{graphicx}
\usepackage{xcolor}
\usepackage{color}
\usepackage{amsmath}
\usepackage{babel}
\usepackage[titletoc]{appendix}
\usepackage{hyperref}
\hypersetup{
     colorlinks = true,
     citecolor  = magenta,  
     linkcolor  = magenta 
}
\usepackage{float}

\newcommand{\beginsupplement}{%
        \setcounter{table}{0}
        \renewcommand{\thetable}{S\arabic{table}}%
        \setcounter{figure}{0}
        \renewcommand{\thefigure}{S\arabic{figure}}%
     }

\begin{document}

\title{Large Fermi-arc and robust Weyl semimetal phase in Ag$_2$S}

\author{Zhenwei Wang$^{1}$}
\thanks{These authors contribute equally to this work.}
\author{Kaifa Luo$^{1}$}
\thanks{These authors contribute equally to this work.}
\author{Jianzhou Zhao$^{2,3}$}
\email{jzzhao@swust.edu.cn}
\author{Rui Yu$^{1}$}
\email{yurui@whu.edu.cn}

\affiliation{$^{1}$ School of Physics and Technology, Wuhan University, Wuhan 430072, China}
\affiliation{$^{2}$ Co-Innovation Center for New Energetic Materials, Southwest University of Science and Technology}
\affiliation{$^{3}$ Theoretical Physics and Station Q Zurich, ETH Zurich, 8093 Zurich, Switzerland}

\begin{abstract}
Three-dimensional Dirac and Weyl semimetals have attracted widespread interest in condensed matter physics and material science. Here, based on first-principles calculations and symmetry analysis, we report that Ag$_2$S with P2$_1$2$_1$2$_1$ symmetry is a topological Dirac semimetal in the absence of spin-orbit coupling (SOC).
Every Dirac points are composed of two Weyl points with the same chirality overlapping in the momentum space.
After taking the SOC into consideration, each Dirac point is split into two Weyl points.
The Weyl points with the opposite chirality are far from each other in the momentum space. 
Therefore, the Weyl pairs are not easy to annihilate and robust in the Ag$_2$S compound, which also lead to long Fermi-arcs on material surface.
The robustness of the Weyl points against the strain is discussed.
\end{abstract}

\maketitle

\section{Introduction}
Topological semimetals with nodal-point-type and nodal-line-type band-crossing points have attracted tremendous attention in condensed matter physics and material science
~\cite{burkov_weyl_2011,wan2011topological,xu2011chern,burkov2011topological,young_dirac_2012,wang_dirac_2012,wang_three-dimensional_2013,weng_weyl_2015,huang2015weyl,borisenko_experimental_2014,lv_experimental_2015,zhao_topological_2016,weng_topological_2015,yu_topological_2015,soluyanov2015type,sun2015prediction,ruan2016symmetry,ruan2016ideal,bzduvsek2016nodal,weng2016topological,wang2017hourglass,armitage2018weyl}.
According to the degree of degeneracy at the band-crossing point, the nodal-points can be further classified as Dirac point, Weyl point, 
and the point beyond Dirac and Weyl types~\cite{Weng_coexistence_2016,wang_prediction_2017,bradlyn_beyond_2016,zhu_triple_2016}.
For example, the Dirac points are four-fold degenerate band-crossing points, which are protected by crystalline symmetries as discussed in references~\cite{yang2014classification,gibson2015three,tang_dirac_2016}.
The Weyl points do not need the protection of crystal symmetries except the translation invariance, 
while the time-reversal or spatial-inversion symmetry must be broken to guarantee their two-fold degeneracy.
These novel band-crossing points in practical materials lead to many exotic transport and optical properties
~\cite{hosur2013recent,wei_excitonic_2012,son_chiral_2013,liu_chiral_2013,ashby_magneto-optical_2013,xu_discovery_2015,xiong_evidence_2015,huang_observation_2015}.
An important hallmark of topological semimetals is their 
unusual surface states, such as the Fermi-arc states for Weyl 
semimetals and the drumhead-like states for nodal-line semimetals
~\cite{essin_bulk-boundary_2011,wan2011topological,xu2011chern,weng_topological_2015,yu_topological_2015}.
The topological nature of the Weyl points can be characterized by their topological charge. 
According to the no-go theorem, the Weyl points with opposite topological charge always appear in pairs
in order to make the total topological charge neutral in the whole Brillouin zone (BZ)~\cite{nielsen1983adler}.
The Weyl points can only be removed if pairs of opposite Weyl points meet and annihilate each other.
A large k-space separation of the Weyl nodes can guarantee a robust Weyl semimetal state, which is a prerequisite for observing the many exotic phenomena in spectroscopic and transport experiments.
Therefore, it is significant to discover more stable, non-toxic and earth-abundant Weyl semimetal~\cite{chang2016strongly,bruno2016observation}, whose Weyl nodes with opposite chirality are separated tremendously in momentum space and locate near the chemical potential in energy.

In this work, we report the Dirac and Weyl states in Ag$_2$S compound.
By means of first-principles calculations and symmetry analysis, we reveal that the SOC splits each Dirac point into two Weyl points with the same topological charge in Ag$_2$S.
The formation of the Weyl points in Ag$_2$S is different from the Weyl points originated from the nodal-line structure. 
For example, in the TaAs family of materials\cite{weng_weyl_2015,huang2015weyl}, the bands inversion leads to 12 nodal-lines in the BZ without SOC. 
After including the SOC effect, each nodal-line is broken but leaves a pair of discrete Weyl points.
The pair of Weyl points originated from the nodal-line have opposite chirality, and the distance between them is dependent on the strength of the SOC.
In the TaAs compound, the distance between each positive-negative charged Weyl pairs is short. 
Therefore, the Weyl points are easy to be annihilate by perturbations and the Fermi-arcs link the Weyl points with opposite charge are short.
The Weyl points in Ag$_2$S are originated from the Dirac points.
Without SOC, bands inversion happens in Ag$_2$S and leads to four Dirac points near the Fermi energy.
The Dirac points with opposite charge, $\pm 2$ as calculated in the later section, are separated in a long distance
in the momentum space.
Including the SOC, each $\pm 2$ charged Dirac point is split into two $\pm 1$ Weyl points. 
While the strength of SOC in Ag$_2$S can not make the long separated and opposite charged Weyl points annihilate each other. 
Therefore the Weyl points are robust and have long Fermi-arc states linked with them on the Ag$_2$S surface.

\section{Crystal structure and calculation method}

\begin{figure}[]
\includegraphics[width=1.0\columnwidth]{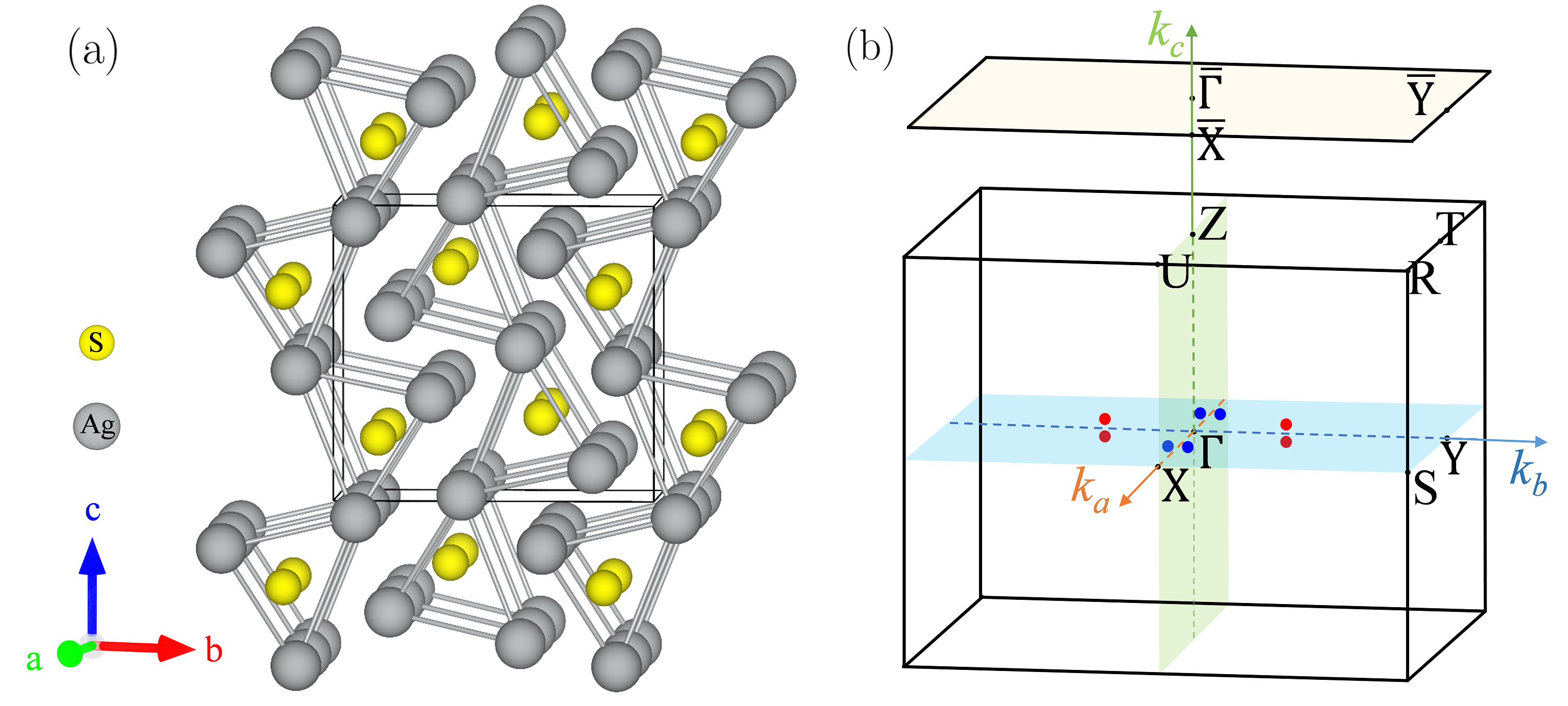}
\caption{
(a) The crystal structure of $P2_12_12_1$ phase Ag$_2$S.
(b) The bulk BZ and its projection onto the (001) direction.
The red (blue) color points indicate the positive (negative) charged Weyl points located on the
$k_x=0$ ($k_z=0$) plane as discussed in section \ref{wsoc}.
}
\label{fig1}
\end{figure}

The Ag$_2$S compound with the nonsymmorphic space group P2$_1$2$_1$2$_1$ (No.~19) is investigated in this work.
The crystal structure is composed of chains of trigonal prisms formed by Ag atoms connected by common edges, each of these prisms being centered by a S atom as shown in 
Fig.~\ref{fig1}(a).
The lattice parameters and the atomic positions of Ag$_2$S was experimentally 
determined~\cite{santamaria-perez_compression_2012} and listed in 
table~\ref{tab:latticeparas}.
\begin{table}[h]
\begin{centering}
\caption{Lattice parameters and atom positions of Ag$_2$S}\label{tab:latticeparas}
\begin{tabular}{cccccccc}
\hline
a~($\AA$) & b~($\AA$) 		 & c~($\AA$) & $\alpha$  & $\beta$    & $\gamma$ \\
6.72500   & 4.14790          & 7.29450   & 90        & 90         & 90 \\
\hline
\hline
  & Site & Wyckoff symbol    & x         & y         & z \\
\hline
1 &  Ag1 &   4a              & 0.02850   & 0.23030   & 0.40790 \\
2 &  Ag2 &   4a              & 0.12740   & 0.40740   & 0.82100 \\
3 &  S 	 &   4a              & 0.22100   & 0.15700   & 0.14500 \\
\hline
\end{tabular}
\par\end{centering}
\end{table}
The symmetry operations of Ag$_2$S crystal include three screw rotations around the principal axes:
$C_{2z}:(x,y,z) \rightarrow ( x+\frac{1}{2}, -y+\frac{1}{2}, -z)$,
$C_{2y}:(x,y,z) \rightarrow (-x, y+\frac{1}{2}, -z)$ and
$C_{2x}:(x,y,z) \rightarrow (-x+\frac{1}{2}, -y, z+\frac{1}{2})$.

The first-principles calculations are performed by using the Vienna {\it ab initio} simulation package (VASP)
~\cite{kresse_efficiency_1996,kresse_efficient_1996} 
based on the generalized gradient approximation (GGA) in the Perdew-Burke-Ernzerhof (PBE) functional and the projector augmented-wave (PAW) pseudo-potential
~\cite{blochl_projector_1994,perdew_generalized_1996}. 
The energy cutoff is set to 400 eV for the plane-wave basis and BZ integration was performed on a regular mesh with 
9$\times$11$\times$7 $\bm k$ points
~\cite{monkhorst_special_1976}. 
The band structure here is also checked by nonlocal Heyd-Scuseria-Ernzerhof (HSE06) hybrid functional calculations
~\cite{heyd_hybrid_2003}.
The surface states are studied by constructing the maximally localized Wannier functions
~\cite{marzari_maximally_1997,souza_maximally_2001,marzari2012maximally} and using the 
WannierTools package~\cite{wu_wanniertools_2018}.

\section{band structure without SOC}\label{wosoc}

\begin{figure}[b]
\includegraphics[width=1.0\columnwidth]{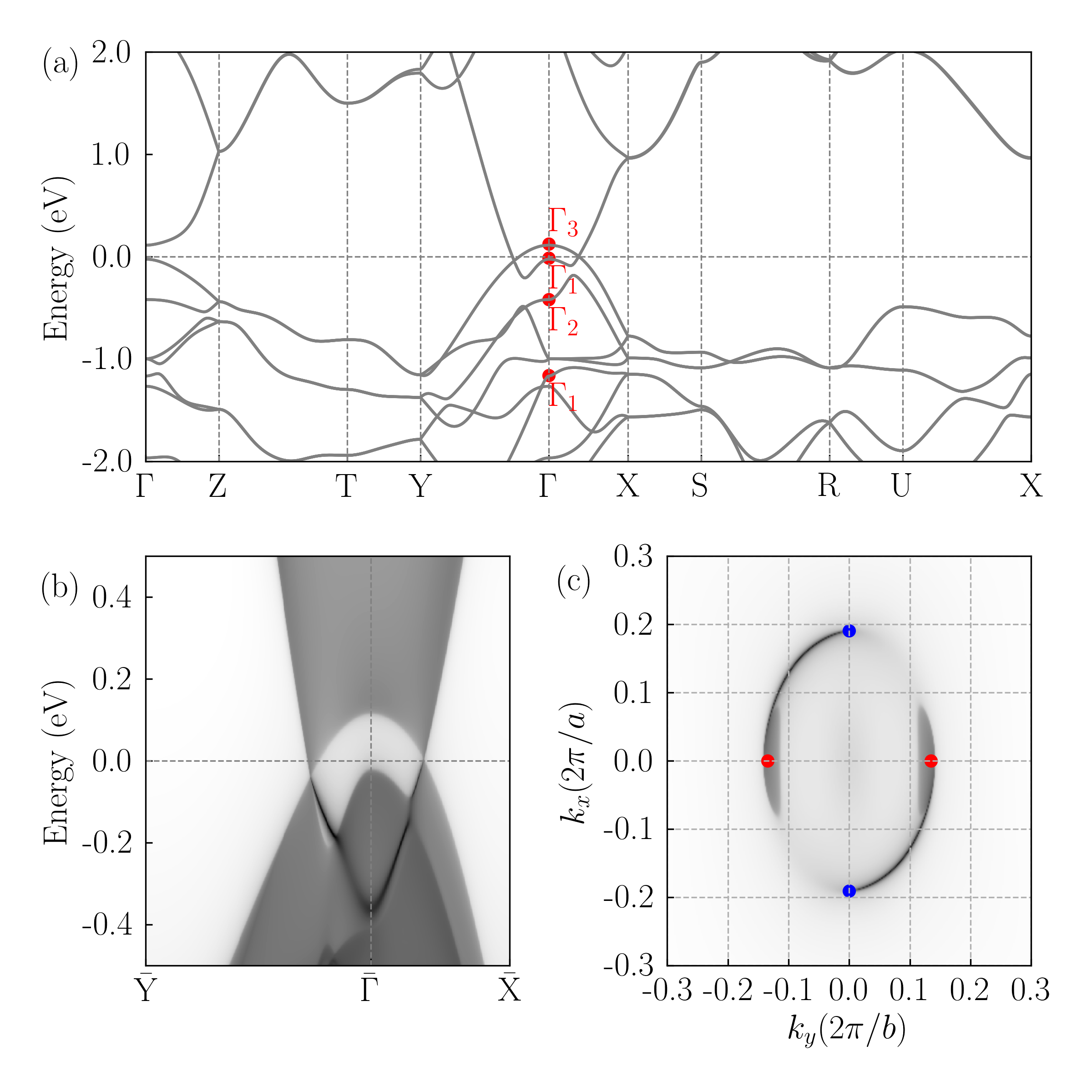}
\caption{
(a) The band structure of $P2_12_12_1$ Ag$_2$S without SOC.
(b) The surface states on the (001) surface.
(c) The Fermi arcs connection the $-2$ Dirac points (blue points) and the $+2$ Dirac points (red points) on the (001) surface at the Fermi energy.
}
\label{fig2}
\end{figure}

The electron configurations for Ag and S atoms are $[Kr]4d^{10}5s^1$ and $[Ne]3s^{2}3p^4$, respectively.
In the Ag$_2$S compound, the Ag atoms have a tendency to lose 5s electrons, while S atoms have a tendency to gain electrons from Ag atoms to form full outer shell.
One may expect an insulating phase for this compounds.
However, the first-principle calculations indicate that Ag$_2$S is semimetal as shown in Fig.~\ref{fig2}(a).
Near the Fermi energy, the valence bands are mostly form S-3p and Ag-4d orbitals, while the conduction band with very strong dispersion is mostly form Ag-5s states.
At the $\Gamma$ point, the energy of Ag-5s band is lower than the S-3p and Ag-4d bands about 1.1 eV, which forms an energy inverted band structure.
The band-crossing points in the $\Gamma$-$X$ and $\Gamma$-$Y$ directions are clearly shown in Fig.~\ref{fig2}(a). 
These band-crossing points are protected by crystalline symmetries, which are revealed by the $\bm k\cdot \bm p$ effective model Eq.~(\ref{eq:H_woS}) near the $\Gamma$ point.
To construct the $\bm k\cdot \bm p$ model Hamiltonian, we find the symmetries at the $\Gamma$ point include the
time reversal symmetry $\hat{T}$ and the $D_2$ point group. The character table for $D_2$ is shown in table~\ref{tab:D2}.
The basis wave-functions for the $\bm k\cdot \bm p$ Hamiltonian are
chosen as the three valence states near the Fermi energy 
and the energy inverted Ag-s states as indicate in Fig.~\ref{fig2}(a).
The irreducible representation for these four states are
$\Gamma_1$, $\Gamma_2$, $\Gamma_1$, and $\Gamma_3$ in order of energy form low to high, which indicates these four wave-functions behave as $s^*$, $p_y$, $s$ and $p_z$ orbitals under the operations of the $D_2$ group symmetry, respectively. 
Here, the $s^*$ orbital is a short notation for Ag-s orbital.

\begin{table}
\caption{Character table and basis functions for point group $D_{2}$. }\label{tab:D2}
\begin{tabular}{cccccc}
\hline
$D_{2}$ & $E$ & $C_{2z}$ & $C_{2y}$ & $C_{2x}$ & linear functions\\
\hline
$\Gamma_{1}$ & +1  &  +1      & +1 & +1 & - \\
$\Gamma_{2}$ & +1  &  -1      & +1 & -1 & y \\
$\Gamma_{3}$ & +1  &  +1      & -1 & -1 & z \\
$\Gamma_{4}$ & +1  &  -1      & -1 & +1 & x \\
\hline
\end{tabular}
\end{table}

The Hamiltonian, projected onto these four bases, that is invariant under all symmetry operations at $\Gamma$ point, has the following form
\begin{widetext}
\begin{equation}
H(\bm k)=
\left(
\begin{array}{cccc}
\epsilon_1 & ia_{12}k_y+b_{12}k_xk_z & a_{13} & a_{14}k_z+ib_{14}k_xk_y+c_{14}k_x^2k_z\\
  & \epsilon_2 & ia_{23}k_y & a_{24}k_x+ib_{24}k_yk_z\\
  &   & \epsilon_3 & a_{34}k_z+ib_{34}k_xk_y\\
\dagger  &   &   & \epsilon_4
\end{array}\label{eq:H_woS}
\right)
\end{equation}
\end{widetext}
up to the third order of $\bm k$. 
It is easy to check that the $s^*$ band and $p_z$ band are decoupled on the $k_x$ and $k_y$ axis, and form band crossing points in these two axises near the Fermi energy.
While they are coupled by $h_{14}=a_{14}k_z$ term, hence there is no band-crossing points along $k_z$ axis.
The derivation of the $\bm k\cdot \bm p$ Hamiltonian, the parameters, and the dispersions form the $\bm k\cdot \bm p$ model compared with the first-principles results are given in the supplementary material.

In the process of constructing the model Hamiltonian Eq.~(\ref{eq:H_woS}), we do not take the SOC into consideration. The spin up and spin down channels are decoupled. We can deal with each channels separately.
Here, we first consider the spin up channel.
The band-crossing points between $s^*$ and $p_z$ states are all twofold degenerate Weyl points.
The Weyl points on the $+k_x$ axis can be rotated to the $-k_x$ direction by $C_{2z}$ operation. 
The $C_{2z}$ rotation does not change the topological charge of the Weyl points, therefore the two Weyl points in $\pm k_x$ axis have the same topological charge.
With the same argument, the two Weyl points on $\pm k_y$ axis also have the same topological charge.
Because the total topological charge of the Weyl points in the BZ must be zero, 
the topological charge of the Weyl points on $\pm k_x$ axis must be opposite to that on the $\pm k_y$ axis.
To determine the particular value of the topological charge, we can check the Berry curvature near these Weyl points. 
The result is that the topological charge for the Weyl points on $\pm k_x$ ($\pm k_y$) axis is negative (positive)
as shown in Fig.~\ref{fig2}(c).
For the spin down channel, there are four Weyl nodes located at the same positions in the BZ and with the same chirality as the spin up channel Weyl nodes due to the time-reversal symmetry.
One of the key character of the Weyl semimetal is the existence of Fermi-arcs on the surface of the material. To calculate the surface states, we generate the Wannier-type tight-binding Hamiltonian for Ag$_2$S. The calculated (001) surface states are shown in Fig.~\ref{fig2}(b,c). It clearly shows that the Fermi-arcs start from the projected points of the positive Weyl nodes and end up at the negative nodes.

\section{band structure with SOC}\label{wsoc}

In this section, we consider the SOC effect on the band-crossing points discussed in the previous section.
Recall the conclusions we get in the previous section that the Weyl points overlapped on the $\pm k_x$ 
axis have opposite chirality to the Weyl points on the $\pm k_y$ axis. 
The strength of the SOC can not make the long separated 
opposite charged Weyl points annihilate with another. 
Therefore the Dirac points are not gaped by SOC in Ag$_2$S but be split into two Weyl points 
and shifted away from the $\pm k_x$ and $\pm k_y$ axis.
The possible positions of the split Weyl points can be determined by symmetry considerations in the following way.
\begin{figure}[b]
\includegraphics[width=1.0\columnwidth]{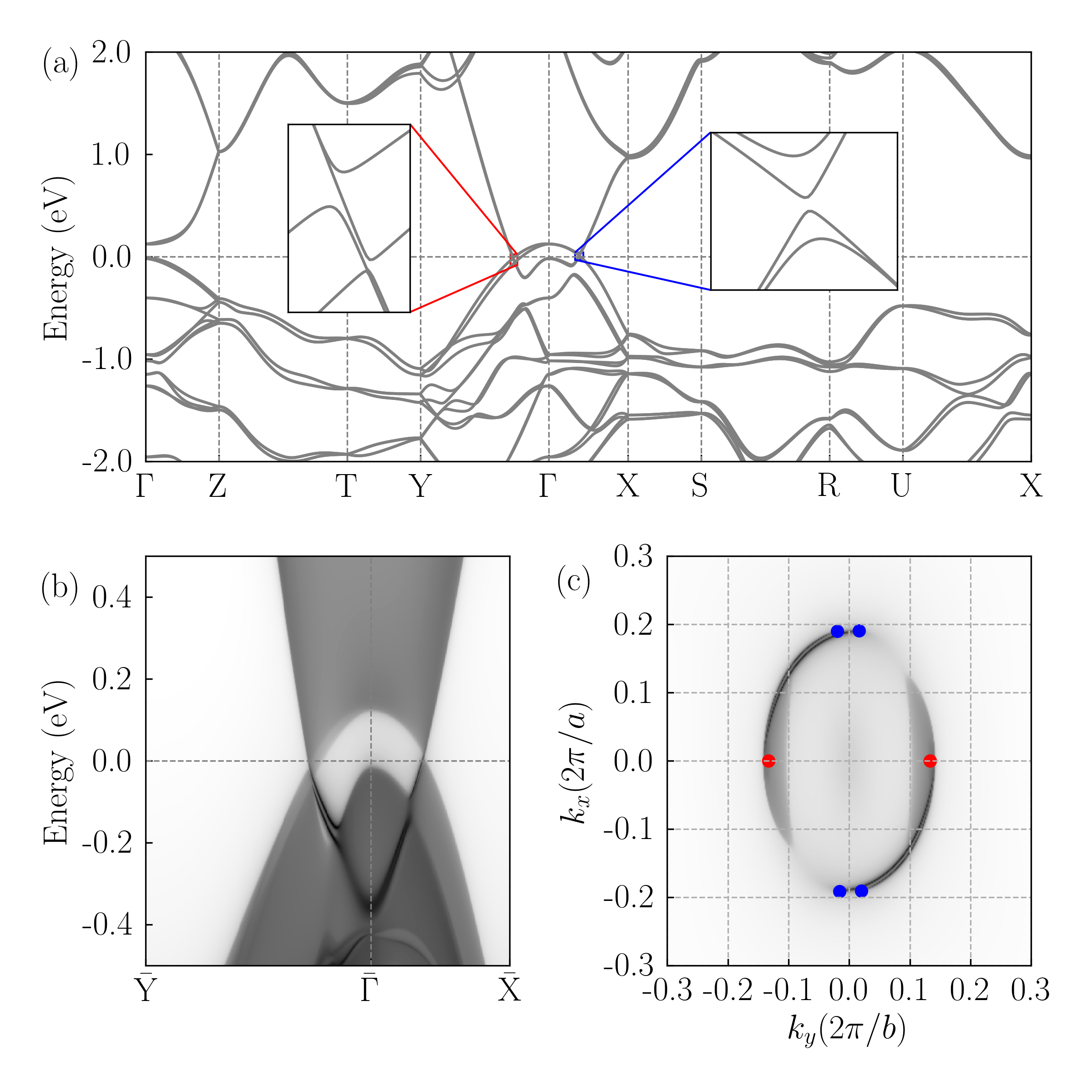}
\caption{
(a) The band structure of Ag$_2$S with SOC. Gap opens around the Dirac points as shown in the inset.
(b) The surface state on the (001) surface.
(c) The Fermi-arcs on the (001) surface with energy located at Fermi-level. Each $-2$ charged Dirac
point is split into two $-1$ Weyl points on the $k_z=0$ plane (blue color points), while each $+2$ charged Dirac point is split into two $+1$ Weyl points on the $k_x=0$ plane (red color points).
}
\label{fig3}
\end{figure}
We denote the position of one of the Weyl points around the $+k_x$ axis as $w_1=(k_x,k_y,k_z)$.
The position of the other one Weyl point can be obtained by $C_{2x}$ rotation, which reads
$w_2=(k_x,-k_y,-k_z)$.
The remain symmetry operators, $\hat C_{2z}$, $\hat C_{2y}$, and $\hat{T}$ transform the above two Weyl points to the $-k_x$ direction:
$w_1\overset{\hat{C}_{2z}}{\rightarrow} (-k_x,-k_y,+k_z)$,
$w_2\overset{\hat{C}_{2z}}{\rightarrow} (-k_x,+k_y,-k_z)$,
$w_1\overset{\hat C_{2y}}{\rightarrow}  (-k_x,+k_y,-k_z)$,
$w_2\overset{\hat C_{2y}}{\rightarrow}  (-k_x,-k_y,+k_z)$,
$w_1\overset{\hat{T}}{\rightarrow}      (-k_x,-k_y,-k_z)$,
$w_2\overset{\hat{T}}{\rightarrow}      (-k_x,+k_y,+k_z)$.
Because there are only two Weyl points around the $-k_x$ axis, which leads to $k_z=0$ or $k_y=0$. 
Therefore, the Weyl points around $\pm k_x$ must located on the $k_z=0$ or $k_y=0$ plane.
With the similar argument, we find that the Weyl points around the $\pm k_y$ axis are shifted to the $k_z=0$ or $k_x=0$ plane.

To check the above results, we perform first-principles calculations with SOC. 
The band-crossing points on the $\pm k_{x,y}$ axises are shifted away as shown in Fig.~\ref{fig3}(a).
Carefully searching the band-crossing points in the whole BZ, 
we get eight Weyl points as schematic shown in Fig.~\ref{fig1}(b).
Four Weyl points with positive chirality located on the $k_x=0$ plane and four with negative chirality located on the $k_z=0$ plane.
Based on the Wannier-type tight-binding Hamiltonian, we compute the (001) surface states as shown in Fig.~\ref{fig3}(b,c).
The projections of the four negative Weyl nodes on $k_z=0$ plane are indicated as blue points.
The positive Weyl points on the $k_x=0$ plane overlap on the 
(001) surface BZ, therefore there are two red points as shown in Fig.~\ref{fig3}(c).
Two Fermi arcs start from the positive Weyl points and end into the negative Weyl points.

\section{Discussions and Conclusions}

\begin{figure}[ ]
\includegraphics[width=1.0\columnwidth]{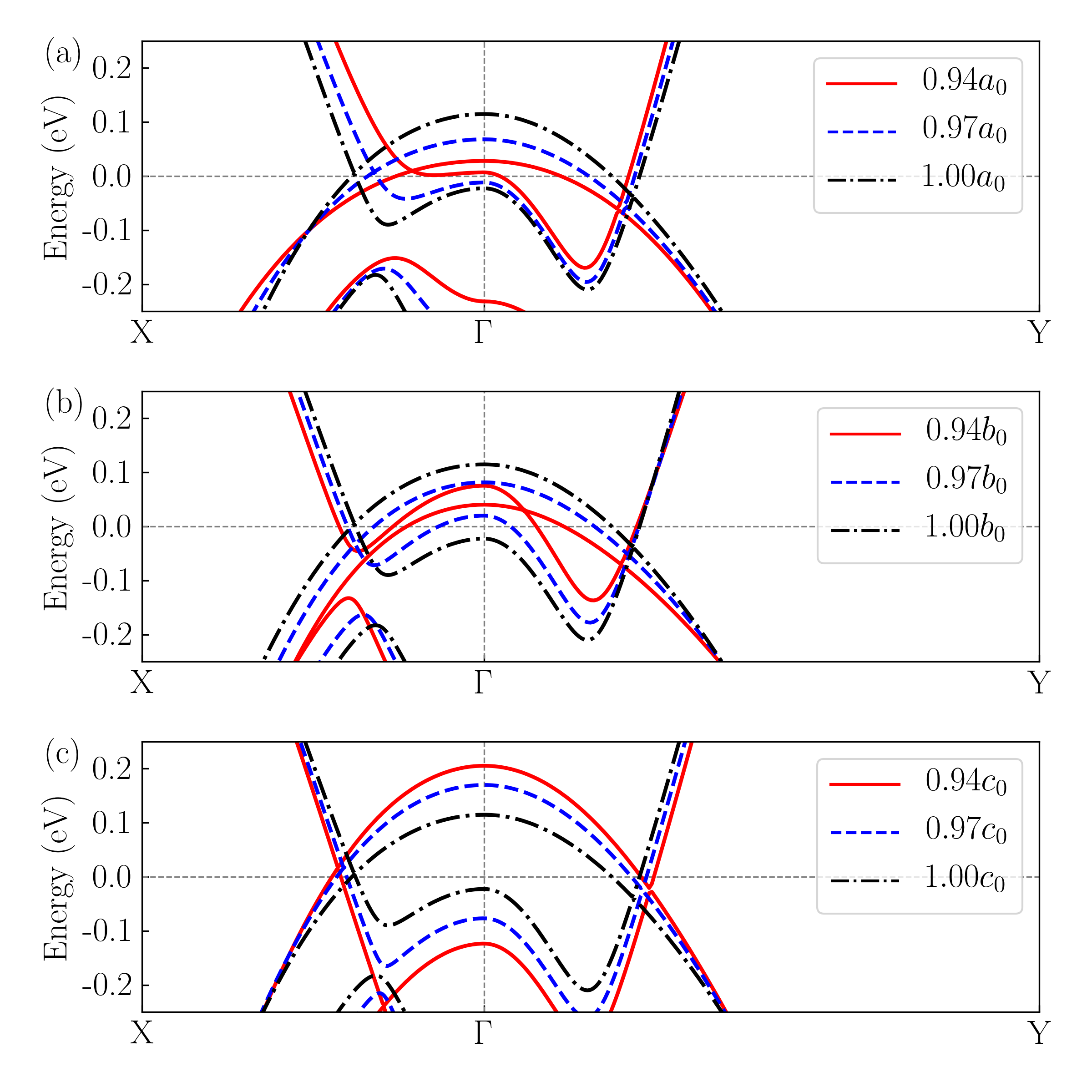}
\caption{
The band structure of Ag$_2$S with $\bm a$, $\bm b$ and $\bm c$ axis compressed by $3\%$ (blue dashed lines) and $6\%$ (red solid lines) compared with uncompressed lattice (black dash-dot lines). 
}
\label{fig4}
\end{figure}

In this section we discuss the stability of the Weyl points against strain in Ag$_2$S.
We perform calculations for adding compressive strain on {\bf a}, {\bf b} and {\bf c} directions, respectively.
The results for $3\%$ and $6\%$ compression of each axises while retain the lattice volume are shown in Fig.~\ref{fig4}.
For the strain adding along $\bm a$ ($\bm b$) axis, the band-crossing points remain for the lattice parameter $a_0$ ($b_0$) is compressed by $6\%$, 
but the distance between the band-crossing point on $k_x$ and $k_y$ direction decrease as the compressive strain increase.
These results indicate that the stability of the Weyl points  is weakened by compressive strain along $\bm a$ and $\bm b$ axises.
For the strain effect adding along $\bm c$ axis, this distance increases as increasing the compressive strain.
Therefore, the stability of the Weyl points is strengthened with the compressive strain along the $\bm c$ axis.

In summary, based on first-principles calculations and symmetry analysis, we predict Ag$_2$S can be tuned from
Dirac semimetal to Weyl semimetal by tuning SOC.
The Dirac points with $\pm 2$ topological charge are protected by the crystalline symmetries and
located on the $k_x$ and $k_y$ axises, respectively. 
The SOC split each Dirac point into a pair of Weyl points with the same chirality, 
while the strength of the SOC can not make the opposite charged Weyl points with large distance in the momentum space meet and annihilate each other.
Therefore, the Weyl points are stable in Ag$_2$S and lead to long  Fermi-arcs on the material surface.

\begin{acknowledgments}
The authors thank Nan Xu for very helpful discussions. 
This work was supported by 
the National Key Research and Development Program of China (No. 2017YFA0304700, No.2017YFA0303402),
the National Natural Science Foundation of China (No. 11674077, No. 11604273).  
J.Z.Z. was also supported by the Longshan academic talent research-supporting program of SWUST (17LZX527), and ETH Zurich funding for his visit.
The numerical calculations in this work have been done on the supercomputing system in the Supercomputing Center of Wuhan University.
\end{acknowledgments}

\section*{Supplementary Material}
\beginsupplement

{\bf The effective model without SOC.}
We first derive the $\bm{k}\cdot\bm{p}$ model Hamiltonian without SOC
near the $\Gamma$ point.
According to the first-principles' results, 
we can chose the three states most near the Fermi energy and one
state with inverted energy at the $\Gamma$ point as basis functions for the $\bm{k}\cdot\bm{p}$ Hamiltonian. 
The irreducible representations for these 
states are $\Gamma_1,~\Gamma_2,~\Gamma_1$, and $\Gamma_3$, 
respectively, in an ascending order of energy. 
These four bands have the symmetries the same as $s,~p_y,~s$, and
 $p_z$ orbitals. 
For convenience, we denote them as 
$\Psi=(\psi_1,\psi_2,\psi_3,\psi_4)^{T}$ hereafter. 

\begin{figure}[H]
    	\centering  
    	\includegraphics[width=0.99\linewidth]{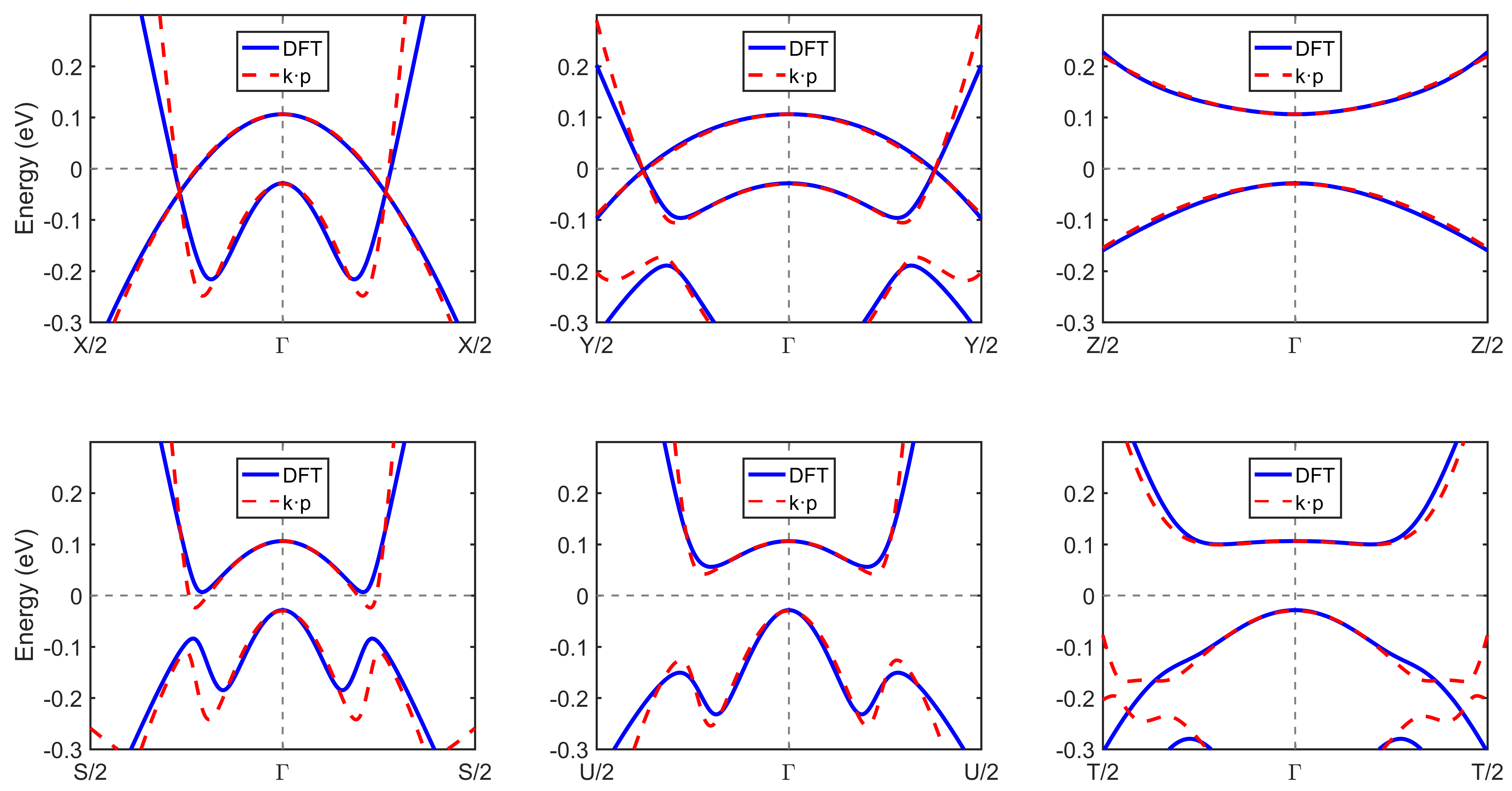} 
    	\caption{The energy dispersion obtained from $\bm{k}\cdot\bm{p}$ model Hamiltonian (red dashed line) is compared with that from the first-principles calculations (blue solid line).}  
    	\label{fig:band_woS} 
    \end{figure}

As stated in the main text, the effective model near the $\Gamma$
point is dictated by time-reversal symmetry $\hat{T}$ and little group $D_2$. 
The model is constructed near the $\Gamma$ point, hence there is no extra phase form the fractional translation for the nonsymmorphic operations.
The $D_2$ point group is generated by two two-fold rotations C$_{2z}$ and C$_{2y}$, which transform our bases as followed:
\begin{equation}
\begin{aligned}
&C_{2y}(\psi_1,\psi_2,\psi_3,\psi_4)^{T}=(\psi_1,\psi_2,\psi_3,-\psi_4)^{T}, \\
&C_{2z}(\psi_1,\psi_2,\psi_3,\psi_4)^{T}=(\psi_1,-\psi_2,\psi_3,\psi_4)^{T}.
\end{aligned}
\end{equation}
Therefore, these symmetry operators can be represented as:
\begin{equation}
\begin{aligned}
& C_{2y}=diag(1,1,1,-1),\\
& C_{2z}=diag(1,-1,1,1),\\
&{T}=C_{2y}\mathcal{K}=diag(1,1,1,-1)\mathcal{K}.
\end{aligned}
\end{equation}
These symmetries give the following constrains to Hamiltonian $H(\bm{k})$:
\begin{equation}
\begin{aligned}
&C_{2y}H(k_x,k_y,k_z)C_{2y}^{-1}=H(-k_{x},k_{y},-k_z),\\
&C_{2z}H(k_x,k_y,k_z)C_{2z}^{-1}=H(-k_{x},-k_{y}, k_z),\\
&{T}   H(k_x,k_y,k_z){T}^{-1}   =H(-k_{x},-k_{y},-k_z).
\end{aligned}\label{cons}
\end{equation}
Keeping up to the third order of $\bm k$, we find the following Hamiltonian 
%
\begin{widetext}

\begin{equation}
H(\bm k)=
\left(
\begin{array}{cccc}
\epsilon_1 & ia_{12}k_y+b_{12}k_xk_z & a_{13} & a_{14}k_z+ib_{14}k_xk_y+c_{14}k_x^2k_z\\
  & \epsilon_2 & ia_{23}k_y & a_{24}k_x+ib_{24}k_yk_z\\
  &   & \epsilon_3 & a_{34}k_z+ib_{34}k_xk_y\\
\dagger  &   &   & \epsilon_4
\end{array}\label{H_woS2}
\right)
\end{equation}
satisfies the constraints in Eq.~(\ref{cons}), where
\begin{equation}
\begin{aligned}
\epsilon_i=\varepsilon_{i0}+\varepsilon_{ij}k_j^2\phantom{00}(i=1,2,3,4,~~j=x,y,z)
\end{aligned}
\end{equation}
\end{widetext}
%
%
and the $a_{ij},~b_{ij},~c_{ij}$ terms denote the first-, second-, third-order couplings between these four orbitals, respectively. 
The parameters in Eq.~(\ref{H_woS2}) are listed in table~\ref{tab:kp_paras}. 
The band dispersions form the $\bm{k}\cdot\bm{p}$ model are
compared with the first-principles' results as shown in Fig.~\ref{fig:band_woS}.

\begin{table}[]
\begin{centering}
\caption{Parameters for the Hamiltonian Eq.~(\ref{H_woS2})}\label{tab:kp_paras}
\begin{tabular}{cccccccc}
\hline
\hline
$\varepsilon_{10}$ &$\varepsilon_{1x}$ & $\varepsilon_{1y}$ & $\varepsilon_{1z}$  & $\varepsilon_{20}$ &$\varepsilon_{2x}$ & $\varepsilon_{2y}$ & $\varepsilon_{2z}$\\
 -0.976 & 49.26 & 10.20 & 3.31  & -0.418 & -8.70 & 11.21 & -1.03\\
\hline
$\varepsilon_{30}$ &$\varepsilon_{3x}$ & $\varepsilon_{3y}$ & $\varepsilon_{3z}$  & $\varepsilon_{40}$ &$\varepsilon_{4x}$ & $\varepsilon_{4y}$ & $\varepsilon_{4z}$\\
 -0.054 & -28.70 & -4.91 & -2.20 & 0.106 & -8.46 & -3.13 & 1.82 \\
\hline
$a_{12}$ & $a_{13}$ & $a_{14}$ & $a_{23}$ & $a_{24}$ & $a_{34}$ & & \\
-0.34 & 0.15 & 0.18 & 0.25 & 0.49 & 0.20  &&\\
\hline
$b_{12}$ & $b_{14}$ & $b_{23}$ & $b_{24}$ & $b_{34}$ & $c_{14}$ &&\\
11 & 4 & 3 & 5 & 6 & 55&&\\
\hline
\hline
\end{tabular}
\par\end{centering}
\end{table}

\bibliography{refs}
\bibliographystyle{apsrev4-1}

\end{document}